\documentclass[12pt]{article}
\usepackage{amsmath}
\usepackage{amssymb}
\usepackage{graphicx}

\begin{document}

\title{Chiral Phenomenological Relations between Rates of Rare Radiative Decay of Kaon to Pion and leptons
and the meson Formfactors.}

\author{V. Pervushin, V. Shilin \\
{\normalsize\it Joint Institute for Nuclear Research},\\
{\normalsize\it 141980, Dubna, Russia.}}
\date{\empty}
\maketitle

\begin{abstract}
In framework of the chiral perturbation theory we obtain the phenomenological relations between
decay branches of rare radiative kaon to pion and
leptons $K^+ \rightarrow \pi^+ l^+ l^-$ and $K^0_S \rightarrow \pi^0 l^+ l^-$ and meson form factors. The
comparison of these results with the present day experimental data shows us that the ChPT relations
for a charge kaon can determine meson form factors from already measured decay rates at high precision
level. However, in the case of the neutral kaon decays $K^0\to \pi^0 e^+e^-(\mu^+\mu^- )$ the
formfactor data are known to a high precision than data on the differential rates of radiative
kaon decay $K^0\to \pi^0 e^+e^-(\mu^+\mu^- )$.
\end{abstract}

\section{Introduction}

New data of decay branches ${\rm
Br}(K^+ \rightarrow \pi^+ l^+ l^-)$ and ${\rm Br}(K^0_S \rightarrow
\pi^0 l^+ l^-)$ were obtained a few years ago in the NA48 experiment
\cite{na48ple, na48ze, na48zmu}. In
analysis of these data a number of theoretical models was used
\cite{eckermain, ambrosioused, friotused, main1, main2}. One of them
is chiral perturbation theory with weak static interactions
\cite{main1, main2} which take into account fermion loops. In this paper, we upgrade this
result in order to study the relation between the decay branches and form factors.

In transitions $K^+ \rightarrow \pi^+ l^+ l^-$ and $K^0_S \rightarrow \pi^0 l^+ l^-$ the main role
is played by one virtual photon exchange: $K \rightarrow \pi\gamma^* \rightarrow \pi l^+ l^-$. To
describe it, we must use the theory of
strong interactions (QCD) and the electroweak theory. Instead of QCD we  use chiral perturbation
theory (ChPT) supposing a contribution of baryon loops in form factors
\cite{main1,main2,pervushdetail,pervushbook}.
To apply electroweak interactions, we  use ChPT in bosonization form and take into account
the meson electromagnetic form factors and their resonance nature.

Main difference of the present paper from other approaches (for
example \cite{eckermain, ambrosioused, friotused, pervushdetail, gershteinkhlopov76}) is
that we have only one coupling constant ($g_8$). Nevertheless, if we take experimentally determined charge radii of
mesons and resonances, our prediction becomes more accuracy. We can conclude that the chiral effective
Lagrangian approach help us to obtain the set of relations between experimental form factors and  decay  branches.

In this article, we ameliorate amplitudes from \cite{main1, main2}, calculate the corresponding decay rates
and test them with available experiments.

\section{Chiral bosonization of EW model}
We start with Lagrangian of weak interactions in bosonized
form \cite{main1}:
$$
\mathcal{L}=-\frac{e}{2\sqrt{2}\sin\theta_{W}}(J^{-}_{\mu}W^{+}_{\mu}+J^{+}_{\mu}W^{-}_{\mu}),
$$
$$
J^{\pm}_{\mu}=[J^1_{\mu}{\pm}iJ^2_{\mu}]\cos\theta_{C}\,+[J^4_{\mu}{\pm}iJ^5_{\mu}]\,\sin\theta_{C},
$$
where Cabbibo angle $\sin\theta_{C}=0.223$.
Using the Gell-Mann matrices $\lambda^k$ one can define the meson current as \cite{pervushbook}:
$$
i\sum\limits \lambda^k J^k_{\mu}=i\lambda^k(V^{k}_{\mu}-A^{k}_{\mu})^{k}=
F^2_\pi e^{i\xi}\partial_\mu e^{-i\xi},
$$
$$
\xi=F_\pi^{-1}\sum\limits_{k=1}^{8}M^k\lambda^k=F_\pi^{-1}\left(
\begin{array}{ccc}
\pi^0+\dfrac{\eta}{\sqrt{3}} & \pi^+\sqrt{2} & K^+\sqrt{2} \\
\pi^-\sqrt{2}  & -\pi^0+\dfrac{\eta}{\sqrt{3}} & K^0\sqrt{2} \\
K^-\sqrt{2} & \overline{K}^0\sqrt{2} & -\dfrac{2\eta}{\sqrt{3}} \\
\end{array}
\right),
$$
here $F_{\pi}\simeq 92.4 \; MeV$. In the first orders in mesons one can write
$$
V^{-}_{\mu}=\sqrt{2}(\sin\theta_{C}\, (K^{-}\partial_{\mu}\pi^{0}-\pi^{0}\partial_{\mu}K^{-} )\,
+\cos\theta_{C}\,(\pi^{-}\partial_{\mu}\pi^{0}-\pi^{0}\partial_{\mu}\pi^{-}))
$$
and
$$
A^{-}_{\mu}=\sqrt{2}\,F_{\pi}\,(\partial_{\mu}K^{-}\sin\theta_{C} + \partial_{\mu}\pi^-\cos\theta_{C}).
$$

This Lagrangian allows us to use the instantaneous weak interaction model \cite{main1, main2}.

\begin{figure}[!htb]
\begin{center}
\includegraphics{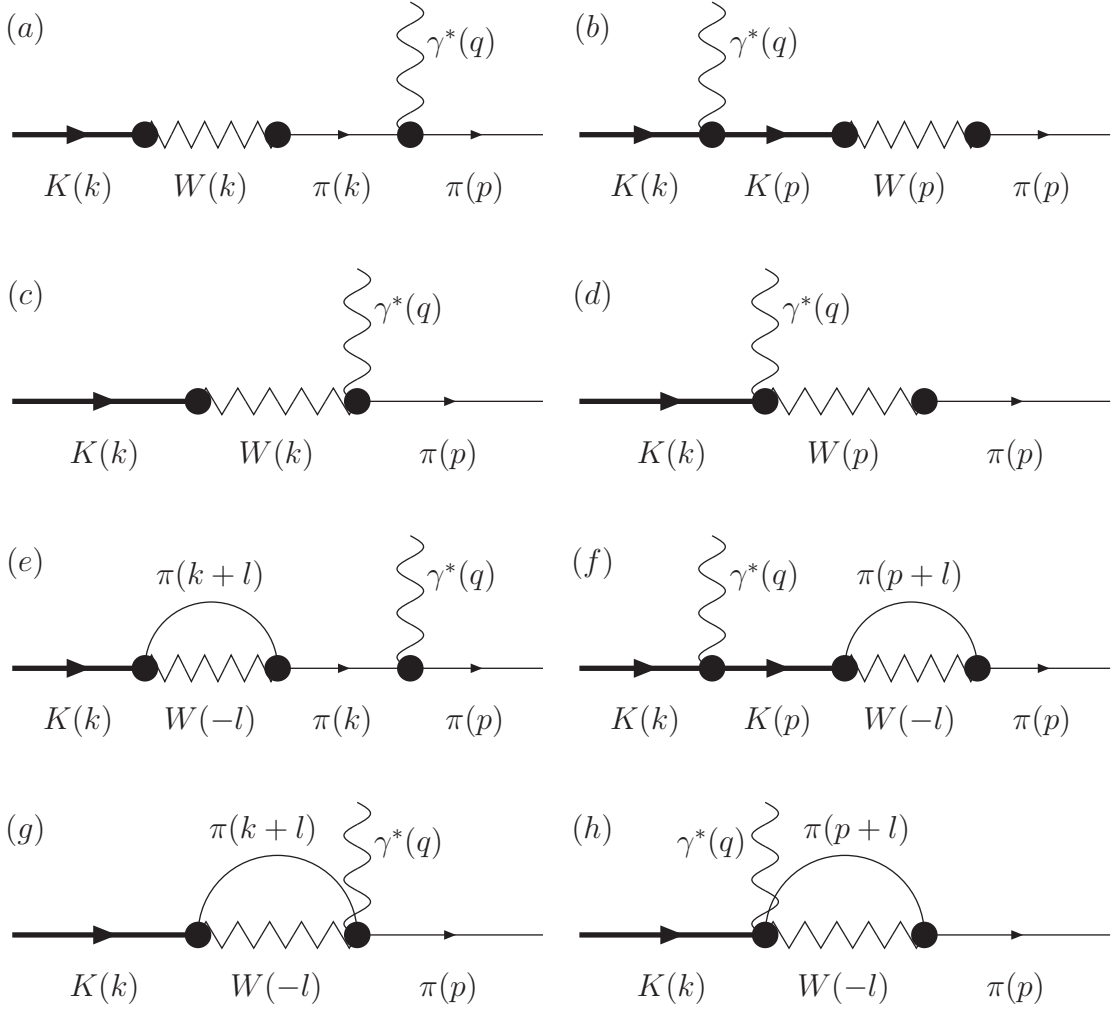}
\caption{Diagrams.}
\label{diagrams}
\end{center}
\end{figure}

\section{The $K \rightarrow \pi l^+ l^-$ amplitude}

In this section we briefly remind the results of the paper \cite{main1}, which we will use in our work. Further
discussions can be find in paper \cite{main2}.

According to \cite{main1, main2}, for the process $K^+ \rightarrow \pi^+ l^+ l^-$ we have diagrams
shown in Fig.\ref{diagrams}, leading to the amplitude:
\begin{equation}
A_{K \rightarrow \pi l^+ l^-} = 2g_8eG_{\rm EW}L_{\nu}D^{\gamma({\rm {rad}})}_{\mu\nu}(q)(k_{\mu }+p_{\mu })
{\cal T}(q^2,k^2,p^2),
\label{amplitudebegin}
\end{equation}
where $g_8 \simeq 5.1$ is the effective parameter of enhancement \cite{main1, main2, eckermain},
$$
G_{\rm EW}= \frac{\sin\theta_{C} \cos\theta_{C}}{8M^{2}_{W}}\frac{e^{2}}{\sin^{2}\theta_{W}}\equiv
\sin\theta_{C}\cos\theta_{C}\frac{G_F}{\sqrt{2}},
$$
$L_{\mu}=\bar{l}\gamma_{\mu}l$ is leptonic current and
\begin{equation}
{\cal T}(q^2,k^2,p^2)=F^2_{\pi}\left(\frac{f^{V}_{\pi}(q^{2})k^{2}}{m^{2}_{\pi}-k^{2}-i\epsilon}
+\frac{f^{V}_{K}(q^{2})p^{2}}{M^{2}_{K}-p^{2}-i\epsilon}
+\frac{f^{A}_{K}(q^{2})+f^{A}_{\pi}(q^{2})}{2}\right)
\label{amplitudedepending}
\end{equation}
Here $F_{\pi} \simeq 92.4 \; MeV$, $f^{V}_{\pi,K}(q^{2})$ and $f^{A}_{\pi,K}(q^{2})$ are
phenomenological meson form factors denoted by fat dots
in Fig.\ref{diagrams} (a, b, e, f) and (c, d, g, h), respectively.

On the mass shell the sum (\ref{amplitudedepending}) takes the form:
$$
{\cal T}(q^2)=F^2_\pi  \left(\frac{f^{A}_{K}(q^{2})+f^{A}_{\pi}(q^{2})}{2}
- f^{V}_{\pi}(q^{2})+ \left(f^{V}_{K}(q^{2})- f^{V}_{\pi}(q^{2})\right)\frac{m_\pi^2}{M_K^2-m_\pi^2}\right).
$$

In case of $K^0_S \rightarrow \pi^0 l^+ l^-$ there are not diagrams Fig.\ref{diagrams} (a - d) and
in the amplitude (\ref{amplitudebegin}) instead of $g_8$ should be $(g_8 -1)$ \cite{main1}.

These amplitudes leads to the decay rate \cite{eckermain, main2, pervushdetail}
\begin{equation}
\Gamma=\bar{\Gamma}_{K \rightarrow \pi l^+ l^-}
\int\limits_{4m^2_l}^{(M_K-m_\pi)^2}{\frac{d q^2}{M_K^2} \rho(q^2)}|\hat{\phi}(q^2)|^2,
\label{decayrate}
\end{equation}
where \cite{eckermain}
$$
\begin{array}{c}
\rho(q^2)=\left(1-\frac{4m_l^2}{q^2}\right)^{1/2}\left(1+\frac{2m_l^2}{q^2}\right)
\lambda^{3/2}\left(1,\frac{q^2}{M^2_K},\frac{m^2_\pi}{M^2_K}\right), \\ [0.5cm]
\lambda(a,b,c)=a^2+b^2+c^2-2(ab+bc+ca), \\ [0.3cm]
\bar{\Gamma}_{K^+ \rightarrow \pi^+ l^+ l^-}=1.37\cdot10^{-19} \; MeV,
\end{array}
$$
and \cite{main1}
$$
\begin{array}{c}
\bar{\Gamma}_{K^0 \rightarrow \pi^0 l^+ l^-}=\left(\frac{g_8-1}{g_8}\right)^2
\cdot\bar{\Gamma}_{K^+ \rightarrow \pi^+ l^+ l^-} \\ [0.5cm]
\end{array}
$$
\begin{equation}
\begin{array}{c}
\hat{\phi}(q^2) = \dfrac{(4\pi)^2{\cal T}(q^2)}{q^2}= \\ [0.5cm]
=\dfrac{(4 \pi F^2_\pi)^2}{q^2} \left(\dfrac{f^{A}_{K}(q^{2})+f^{A}_{\pi}(q^{2})}{2}
- f^{V}_{\pi}(q^{2})+ \left(f^{V}_{K}(q^{2})- f^{V}_{\pi}(q^{2})\right)\dfrac{m_\pi^2}{M_K^2-m_\pi^2}\right)
\end{array}
\label{phi}
\end{equation}
Thus, ChPT and instantaneous weak interaction model leads to formulas (\ref{decayrate}) and (\ref{phi}) as
relationship between decay rates and formfactors.

\section{Form factors}
\subsection{$K^+ \rightarrow \pi^+ l^+ l^-$.}
One can make an assumption that electromagnetic form factors of the kaon and pion are saturated with resonances as
in the $\rho$-dominance model. One of possible models of such the suturation is ChPT with both meson and baryon
loops \cite{pervushbook, belkov1995, belkov2005, belkovrule}, so in \cite{main1, main2} at small $q^2$ they
were chosen in the form
\begin{equation}
\begin{split}
f_{\pi}^{V}(q^2)\simeq f_K^{V}(q^2)\simeq f^{V}(q^2)&=1+M^{-2}_\rho q^2+\alpha_{0}\Pi_\pi(q^2)+\ldots , \\
f_{\pi}^{A}(q^2)\simeq f_K^{A}(q^2)\simeq f^{A}(q^2)&=1+M^{-2}_{{a_0}^1} q^2+\ldots.
\end{split}
\label{formfactororiginal}
\end{equation}
We can calculate decay rates using the  resonances \cite{pdg}:
\begin{equation}
\begin{array}{ll}
M_\rho = 775.49 \pm 0.34 \; MeV, & I^G(J^{PC}) = 1^+(1^{--}), \\
M_{{a_0}^1} = 980 \pm 20 \; MeV, & I^G(J^{PC}) = 1^-(0^{++});
\end{array}
\label{mrhoma}
\end{equation}
and pion loop contribution:
$$
\alpha_0=\frac{4}{3}\frac{m_{\pi^+}^2}{(4\pi F_\pi)^2}=0.01926 \pm 0.00077,
$$
\begin{equation}
\begin{array}{ll}
\Pi_\pi(t)=(1-\bar{t})\left(\dfrac{1}{\bar{t}}-1\right)^{1/2}\arctan\left(\dfrac{\bar{t}^{1/2}}{(1-\bar{t})^{1/2}}\right)-1,
& \bar{t}=\dfrac{t}{(2m_{\pi^+})^2}<1; \\
\Pi_\pi(t)=\dfrac{\bar{t}-1}{2}\left(1-\dfrac{1}{\bar{t}}\right)^{1/2}
\left\{i\pi- \log \dfrac{\bar{t}^{1/2}+(\bar{t}-1)^{1/2}}{\bar{t}^{1/2}-(\bar{t}-1)^{1/2}}\right\}-1,
& \bar{t}\geqslant 1.
\end{array}
\label{pionloop}
\end{equation}

Let us make two remarks at this point.

First, $f_{\pi}^{V}(q^2)$ and $f_K^{V}(q^2)$ are nothing but electromagnetic form factors of the charged pion
and kaon, but we know them much better from experiment\cite{pdg}. So we can prove $f^V(q^2)$ using experimental
data. At $q^2\rightarrow 0$:
\begin{equation}
f^V(q^2\rightarrow 0)\simeq 1+ \frac{< \! r^2 \! >}{6(\hbar c)^2} \, q^2
\label{formfactornearzero}
\end{equation}
\begin{equation}
\begin{array}{l}
{<\! r \!>}_{\pi^+}=0.672 \pm 0.008 \; fm, \\
{<\! r \!>}_{K^+}=0.560 \pm 0.031 \; fm.
\end{array}
\label{radiusplus}
\end{equation}
Of course, in $< \! r \! >$ the $\Pi_\pi(q^2)$ term is already included. To retrieve  $\Pi_\pi(q^2)$ (and
nontrivial $q^2$-dependence), expand it in series near zero:
\begin{equation}
\alpha_0\Pi_\pi(q^2\rightarrow 0)\simeq-\alpha_0\frac{4}{3}\frac{q^2}{(2m_{\pi^+})^2},
\label{pionloopnearzero}
\end{equation}
subtract (\ref{pionloopnearzero}) from (\ref{formfactornearzero}) and add (\ref{pionloop}):
\begin{equation}
f^V(q^2\rightarrow 0)\simeq 1+ \left( \dfrac{{<r>}^2}{6(\hbar c)^2} +
\alpha_0\dfrac{4}{3}\dfrac{1}{(2m_{\pi^+})^2}\right)q^2 + \alpha_0 \Pi_\pi(q^2)
\label{formfactorplusnearzero}
\end{equation}

At large $q^2$, $f_{\pi^+}^{V}(q^2)$ and $f_{K^+}^{V}(q^2)$ have maximum at $q^2=M^2_{\rho}$.

Second, beside ${a_0}^1$ there is:
\begin{equation}
M_{{a_0}^2} = 1474 \pm 19 \; MeV, \; I^G(J^{PC}) = 1^-(0^{++}).
\label{ma}
\end{equation}
If ${a_0}^2$ is not taken into account, a huge discrepancy with experiment results will be got.

Finally, using (\ref{mrhoma}), (\ref{pionloop}), (\ref{radiusplus}), (\ref{formfactorplusnearzero}), (\ref{ma}) we
have the following improved hypothesis of (\ref{formfactororiginal}) in Pad\'e type approximations:
\begin{equation}
\begin{array}{l}
f_{\pi^+}^{V}(q^2)=\dfrac{\gamma_\pi}{1-\dfrac{1}{\gamma_{\pi}} \left( \left(
\dfrac{{<r>}_{\pi^+}^2}{6(\hbar c)^2} + \alpha_0\dfrac{4}{3}\dfrac{1}{(2m_{\pi^+})^2}\right)q^2 +
\alpha_0 \Pi_\pi(q^2)\right)}+(1-\gamma_\pi) \\ [1.5cm]
f_{K^+}^{V}(q^2)=\dfrac{\gamma_K}{1-\dfrac{1}{\gamma_K} \left( \left(
\dfrac{{<r>}_{K^+}^2}{6(\hbar c)^2} + \alpha_0\dfrac{4}{3}\dfrac{1}{(2m_{\pi^+})^2}\right)q^2 +
\alpha_0 \Pi_\pi(q^2)\right)}+(1-\gamma_K) \\ [1.5cm]
f_{\pi^+}^{A}(q^2)\simeq f_{K^+}^{A}(q^2)\simeq f^{A}(q^2)=
\dfrac{1}{1-\dfrac{q^2}{M^{2}_{{a_0}^1}}}+\dfrac{1}{1-\dfrac{q^2}{M^{2}_{{a_0}^2}}}-1,
\end{array}
\label{formfactorplus}
\end{equation}
$\gamma_\pi=1.176677$ and $\gamma_K=0.855628$ have been chosen to put the position of maximum
of $f_{\pi^+}^{V}(q^2)$ and $f_{K^+}^{V}(q^2)$ to $q^2=M^2_{\rho}$. At small $q^2$:
$$
f^{A}(q^2)=1+\dfrac{q^2}{M^{2}_{{a_0}^1}}+\dfrac{q^2}{M^{2}_{{a_0}^2}}+\ldots
$$

\begin{figure}[!htb]
\begin{center}
\includegraphics{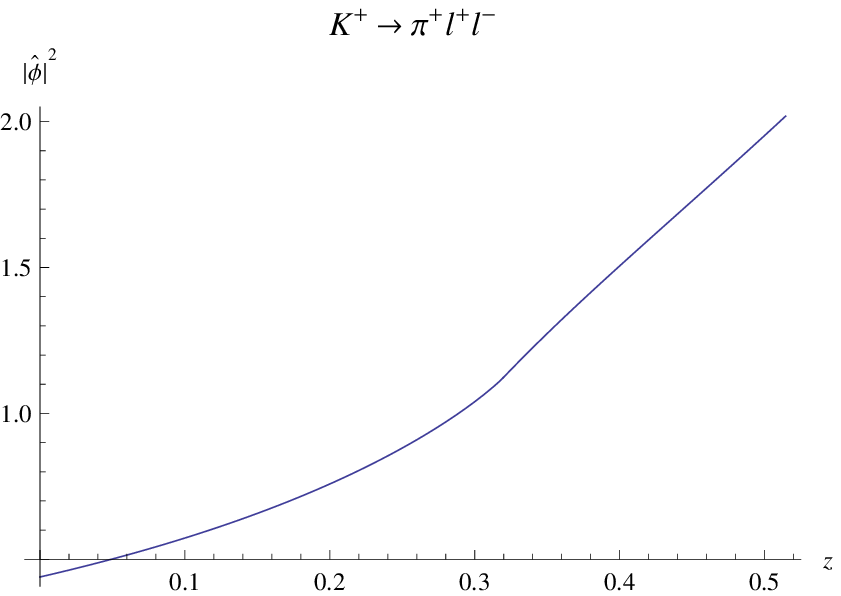}
\caption{The $|\hat{\phi}(q^2)|^2$ defined by (\ref{phi}) and (\ref{formfactorplus}).}
\label{phisquareplus}
\end{center}
\end{figure}

A plot of (\ref{phi}) with (\ref{formfactorplus}) is shown in Fig.\ref{phisquareplus}, $z=\frac{q^2}{M^2_{K^+}}$.

\subsection{$K^0_S \rightarrow \pi^0 l^+ l^-$.}
In this case we have\cite{pdg}:
\begin{equation}
{<\! r^2 \!>}_{K^0}=-0.077 \pm 0.010 \; fm^2,
\label{radiusneutralkaon}
\end{equation}
and we can neglect the neutral pion electromagnetic radius \cite{neutralpionradius}:
\begin{equation}
{<\! r^2 \!>}_{\pi^0}=0.
\label{radiusneutralpion}
\end{equation}

Notice that:
$$
\begin{array}{c}
\dfrac{{<\! r^2 \!>}_{K^0}}{6(\hbar c)^2} \simeq -0.33 \times 10^{-6} \; MeV \\ [0.5cm]
\dfrac{d \alpha_0 \Pi_\pi}{d q^2}(0) \simeq -0.33 \times 10^{-6} \; MeV
\end{array}
$$
which means that  ${<\! r^2 \!>}_{K^0}$ is determined almost only by  $\Pi_\pi(q^2)$, that is why we will not
use resonance behavior of $f_{\pi}^{V}(q^2)$ and $f_K^{V}(q^2)$:
\begin{equation}
\begin{array}{l}
f_{\pi^0}^{V}(q^2)=0 \\ [0.8cm]
f_{K^0}^{V}(q^2)=\left(\dfrac{{<r>}_{K^0}^2}{6(\hbar c)^2} + \alpha_0\dfrac{4}{3}\dfrac{1}{(2m_{\pi^+})^2}\right)q^2+
\alpha_0 \Pi_\pi(q^2) \\ [0.8cm]
f_{\pi^0}^{A}(q^2)\simeq f_{K^0}^{A}(q^2)\simeq f^{A}(q^2)=
\dfrac{1}{1-\dfrac{q^2}{M^{2}_{{a_0}^1}}}+\dfrac{1}{1-\dfrac{q^2}{M^{2}_{{a_0}^2}}}-2
\end{array}
\label{formfactorzero}
\end{equation}
At small $q^2$:
$$
f^{A}(q^2)=\dfrac{q^2}{M^{2}_{{a_0}^1}}+\dfrac{q^2}{M^{2}_{{a_0}^2}}+\ldots
$$

\begin{figure}[!htb]
\begin{center}
\includegraphics{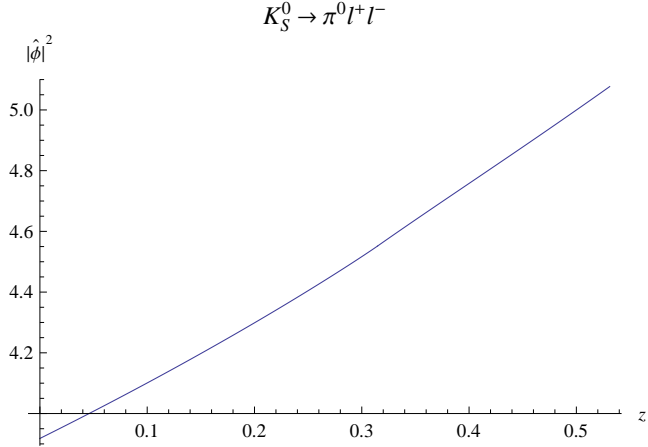}
\caption{The $|\hat{\phi}(q^2)|^2$ defined by (\ref{phi}) and (\ref{formfactorzero}).}
\label{phisquarezero}
\end{center}
\end{figure}

A plot of (\ref{phi}) with (\ref{formfactorzero}) is shown in Fig.\ref{phisquarezero}, $z=\frac{q^2}{M^2_{K^0}}$.

\section{Decay rates}
\begin{figure}[!htb]
\includegraphics[width=0.5\textwidth]{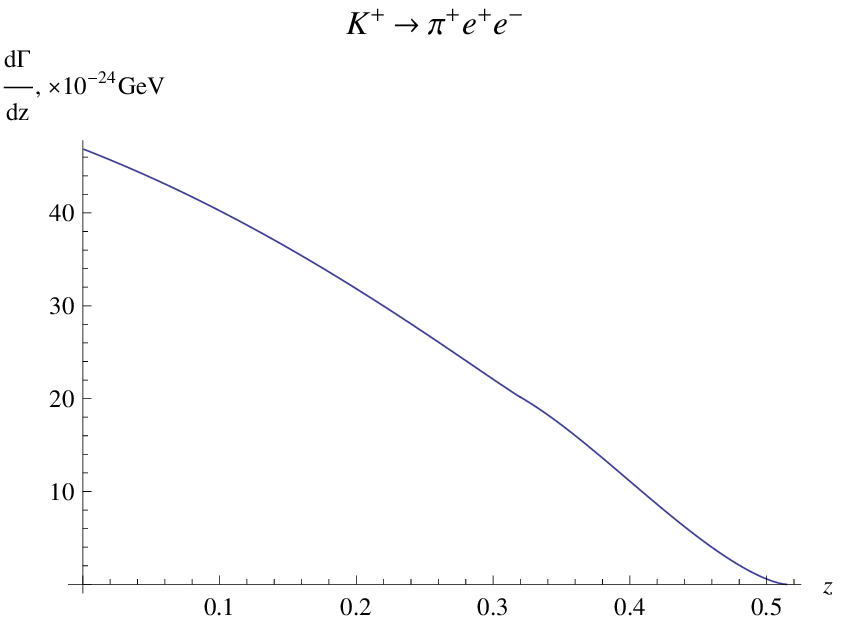}
\includegraphics[width=0.5\textwidth]{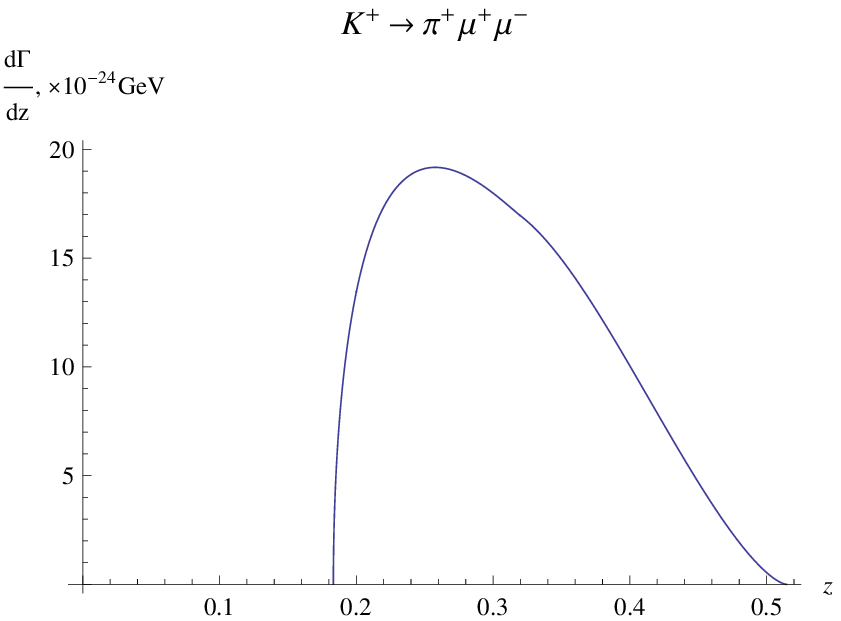}
\includegraphics[width=0.5\textwidth]{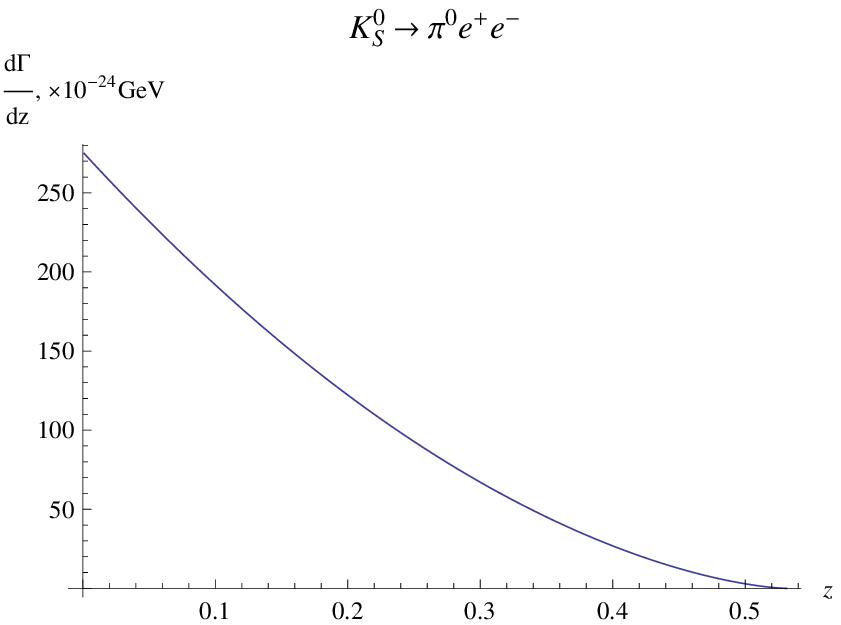}
\includegraphics[width=0.5\textwidth]{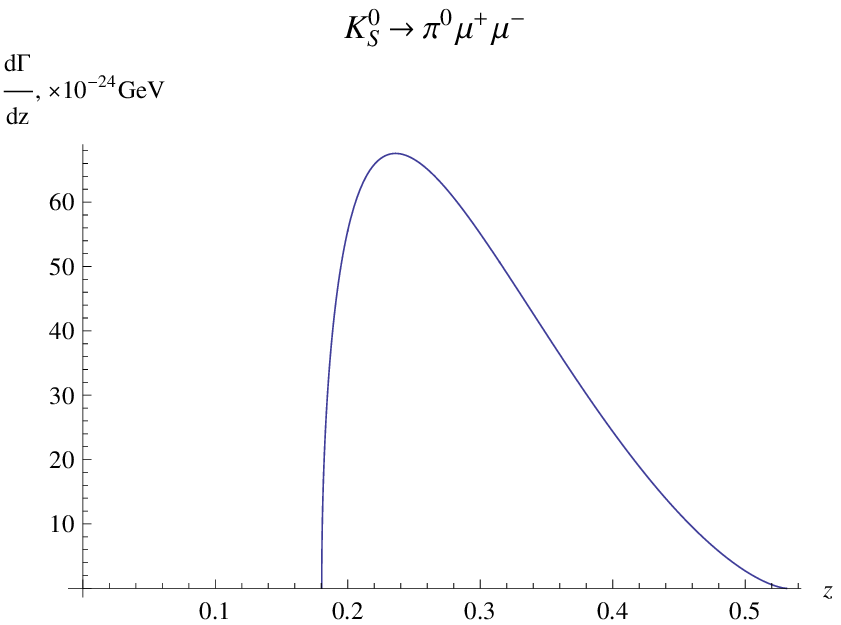}
\caption{The $\frac{d \Gamma}{d z}$ determined by relations (\ref{formfactorplus}),(\ref{formfactorzero}),(\ref{phi}),(\ref{decayrate}).}
\label{diffdecays}
\end{figure}

If we substitute formulae (\ref{formfactorplus}) and (\ref{formfactorzero}) into equations
(\ref{phi}) and (\ref{decayrate}), we get decay rates summarized
in table \ref{decays}. We can see good agreement with experiments in all cases.
Large inaccuracy in $K^+$ decays
arises from subtraction approximately equal to $f^{A}$ and $f_{\pi^+}^{V}$ (\ref{formfactorplus})
in formula (\ref{phi}). This table shows us that at present day precision level, better to
extract $f_{\pi^+}^{V}$ and $f^{A}$ from decay rates $K^+ \rightarrow \pi^+ l^+ l^-$.
\begin{table}[!htb]
\centering
\caption{Decay rates compared with experiments, $MeV$.}
\begin{tabular}{|c|c|c|}\hline
& $\Gamma$ & $\Gamma_{exp}$ \\ \hline
$K^+ \rightarrow \pi^+ e^+ e^- \!\phantom{\dfrac{1}{1}}$ & $(1.29 \pm 0.40)\times 10^{-20}$ & $(1.654 \pm 0.064)\times 10^{-20}$ \cite{na48ple} \\ \hline
$K^+ \rightarrow \pi^+ e^+ e^-, \; z>0.08 \!\phantom{\dfrac{1}{1}}$ & $(0.94 \pm 0.28)\times 10^{-20}$ & $(1.212 \pm 0.043)\times 10^{-20}$ \cite{na48ple} \\ \hline
$K^+ \rightarrow \pi^+ \mu^+ \mu^- \!\phantom{\dfrac{1}{1}}$ & $(0.39 \pm 0.11)\times 10^{-20}$ & $(0.431 \pm 0.075)\times 10^{-20}$ \cite{pdg} \\ \hline
$K^0_S \rightarrow \pi^0 e^+ e^- \!\phantom{\dfrac{1}{1}}$ & $(5.41 \pm 0.68)\times 10^{-20}$ & $(4.3 \begin{array}{l}+2.2\\-1.9\end{array}) \times 10^{-20}$ \cite{na48ze} \\ \hline
$K^0_S \rightarrow \pi^0 e^+ e^-, \; q>165 \!\phantom{\dfrac{1}{1}}$ & $(2.90 \pm 0.37)\times 10^{-20}$ & $(2.2 \begin{array}{l}+1.1\\-0.9\end{array}) \times 10^{-20}$ \cite{na48ze} \\ \hline
$K^0_S \rightarrow \pi^0 \mu^+ \mu^- \!\phantom{\dfrac{1}{1}}$ & $(1.23 \pm 0.16)\times 10^{-20}$ & $(2.1 \begin{array}{l}+1.1\\-0.9\end{array}) \times 10^{-20}$ \cite{na48zmu} \\ \hline
\end{tabular}
\label{decays}
\end{table}
Differential decay rates are presented in Fig.\ref{diffdecays}.

\section{Conclusion}
In framework of ChPT we calculated decay rates
of $K^+ \rightarrow \pi^+ l^+ l^-$ and $K^0_S \rightarrow \pi^0 l^+ l^-$ using
measured electromagnetic meson radii \cite{pdg} and inserting resonances with quantum numbers of
$a_0$-meson into formfactors in the instantaneous weak interaction.
Taking into account the instantaneous weak interaction is the
difference of our approach from other ones.

The results we obtained to be in good agreement with experiments, for instance one can determine the neutral
kaon decay branch data using the meson form factor data. However, there is a large amount
of inaccuracy. On the other hand, the high sensitivity of obtained decay rates allows us for a charge kaon
to determine the form factors and masses of $a_0$ mesons from
already measured $\Gamma(K^+ \rightarrow \pi^+ l^+ l^-)$ at high precision level.


\section*{Acknowledgments}
The authors thank  Profs. M.K.~Volkov,  L. Litov,  V. Kekelidze, and
Yu. S. Surovtsev for  useful discussions. One of us (V.S.) is
grateful to Direction of the Bogoliubov Laboratory of Theoretical
Physics for the hospitality.

\end{document}